# First-Principles Investigation of Anchoring Behavior of $WS_2$ and $WSe_2$ for Lithium-Sulfur Batteries

Rahul Jayan, Md Mahbubul Islam*

Department of Mechanical Engineering, Wayne State University, Detroit, MI 48202

*Corresponding author: mahbub.islam@wayne.edu

The commercial realization of lithium-sulfur (Li-S) batteries is obstructed because of rapid capacity fading due to lithium polysulfides (LiPSs) dissolution into the electrolyte. In order to enhance the efficiency and performance of the Li-S batteries, the transition metal dichalcogenides are reported as promising anchoring materials (AMs) as they could strongly adsorb and effectively suppress the migration of the polysulfides species. Herein, we used first-principles based density functional theory (DFT) calculations to investigate the interactions between AMs such as tungsten dichalcogenides, $WX_2$ (X=S and Se) and the LiPSs. The LiPSs binding behavior of $WS_2$ and $WSe_2$ are found to be quite similar. The calculated adsorption energies of LiPS species indicate that the $WX_2$ possesses moderate binding strength and the binding is facilitated via charge transfer from the polysulfides to the AM. We observe elongation of intramolecular Li-S bonds in LiPS upon their adsorption onto the $WX_2$, however, chemical structures of LiPSs are preserved without decomposition. The calculated density of states indicates the LiPS adsorbed $WX_2$ systems exhibits semiconducting behavior with a slightly lower bandgap compared to the pristine $WX_2$. Overall, our simulation results provide detailed insight into the behavior of $WX_2$ as AMs to suppress the LiPSs migration and henceforth paves the way towards the development of high-performance Li-S batteries.

## 1. Introduction

The ever-increasing energy demand for the electrification of transportation systems and large-scale storage for renewable energies triggered tremendous research thrust towards developing safe and high-performance rechargeable battery systems. The Li-ion batteries, ubiquitously used in portable electric systems, have insufficient capacity to meet the future energy demand. Alternative battery technology such as lithium-sulfur (Li-S) redox couples is considered as the most promising candidate for next-generation portable electronics and electric vehicles owing to its high energy density (2600 Whkg$^{-1}$) and specific capacity (1675 mAh g$^{-1}$). The theoretical energy density and capacity of sulfur cathodes are at least ten times greater than the widely used traditional transition metal oxide cathodes of Li-ion batteries.[1–7] The elemental sulfur is abundant in nature, non-toxic, and cheap; additionally, the lower cell potential of Li-S batteries offers safer operations.[8]

In spite of all the advantages, the commercialization of Li-S batteries has not yet been possible because of several critical issues that adversely affect the cyclic reversibility and rate capability. During the cell discharging process, the elemental sulfur reduces to convert to polysulfides and the dissolution of higher-order polysulfides ($Li_2S_n$, n = 4,6, and 8) into the electrolyte creates shuttling between the electrodes. The further reduction of the larger polysulfides leads to lower-order polysulfides such as $Li_2S$ and $Li_2S_2$, which deposits and passivate the electrode surfaces.[7,9,10] The passivation layer increases cell resistance and causes active mass loss as such is responsible for poor cycling capability and rapid capacity fading. Additionally, significant volume changes of sulfur materials during charging and discharging causes pulverization and mechanical failure of the cathode material.[11,12]

The insulating nature of the elemental sulfur and lithium polysulfides (LiPSs) require to use conductive matrix as a current collector. Traditionally, carbon-based materials have been widely used because of their good electric conductivity and large surface areas.[5,13] However, the apolar carbon host materials offer poor binding to polysulfides and exhibit limited performance towards polysulfides retention within the cathode material. Trapping of the polysulfides during the discharge process is a key factor to inhibit their further dissolution into the electrolyte.[2,9,14] Recently, research efforts have been expended to find anchoring materials (AMs) alternative to the carbon-based materials. The materials with polar characteristics and moderate binding strength with the LiPSs are desirable for the improved performance of the cathode. Modifications of carbon surface by introducing amphiphilic polymers,[15] using polar nanostructured AM such as polymers (polyaniline, polypyrole,

PEDOT)[15–17], metal oxides (TiO$_2$, SiO$_2$, Al$_2$O$_3$)[18–24] are reported to improve polysulfide interactions and to prevent their dissolution.

Furthermore, the distinct layered structure and wide band gap (semiconductor) properties of transition metal dichalcogenides (TMDs) are found appealing for the various energy storage applications,[25] solar cells,[26–28] hydrogen evolution reactions.[26,28–30] Ghazi et al. reported improved polysulfide retention in a MoS$_2$/celgard composite cathode which exhibits a stable capacity of 400 mAhg$^{-1}$ with columbic efficiency of 99% for 600 cycles.[31] A stable capacity of 900 mAhg$^{-1}$ at a current rate of 0.2 C was observed for MoS$_2$/SnO$_2$ composite cathode.[32] Similar to the MoS$_2$, other dichalcogenides such as WS$_2$ has also been studied to realize its performance as AM. For example, Huang et al. studied the reaction mechanism of the LiPSs with the polar WS$_2$ contained in the three dimensional reduced graphene oxide/carbon nanotube aerogel.[33] The WS$_2$ was reported to provide good binding strength and catalytic activity for enhanced reaction kinetics of polysulfides conversions.[33] The edge sites of the WS$_2$ and MoS$_2$ are reported as electrocatalytically active sites for the reversible conversion of the LiPSs.[34] Naresh et al. observed preferential absorption of LiPSs on the WS$_2$ surfaces and reduction of the redox overpotential as well as an increased surface diffusion kinetics of LiPSs.[35]

Apart from the experimental studies, several computational studies are performed to obtain further insights into the mechanisms of AMs to inhibit polysulfide shuttling and to understand the electronic properties of the host materials. Predominantly, transition metal oxides, sulfides, and MXenes are studied using density functional theory (DFT) simulations. The bare MXenes provide strong binding that leads to the decomposition of the polysulfides that hinders the conversion of polysulfides.[36,37] However, the effectiveness of the MXenes is reported to be improved via various functionalization. Sim et al. investigated the anchoring behavior of F and O functionalized Ti$_2$C MXenes with the LiPSs and reported moderate binding energies that can suppress the LiPSs shuttling.[38] Moderate LiPSs adsorption behavior was also demonstrated for V$_2$CS$_2$ MXene.[39] Wang et al. conducted a detailed study on the anchoring behavior of various two-dimensional layered materials (oxides, sulfides, and chlorides) and delineated the mechanisms of LiPSs binding on the AMs.[22] The role of the structural defects on the AMs are also investigated. The defect sites cause strong binding of LiPSs due to an increased charge transfer between the AM and LiPSs.[39] Various transition metal (TM) oxides and sulfides, nitrogen, and metal doped graphene, boron doped graphidyn, and borophene are reported to provide improved adsorption kinetics of the polysulfides.[40–43] It has also been illustrated as the balance between the affinity for and moderate reactivity to sulfur are essential for the effectiveness of AMs.[40] Such advancements yet lack a mechanistic understanding of the interactions between LiPSs and greatly promising[34,35] dichalcogenides of tungsten WX$_2$ (X=S and Se).

Herein, for the first time, we carry out first-principles calculations based on DFT to investigate the adsorption behavior of polysulfides with tungsten chalcogenides WX$_2$ (X=S and Se). We calculate the binding energies of various LiPSs on the WX$_2$ in order to obtain insights into the polysulfide retention capabilities of these materials. We observed the trend of the LiPSs adsorption behavior on WX$_2$ is similar to the MoS$_2$.[22] Our study illustrates that the WX$_2$ provides moderate bindings to prevent polysulfides dissolution without chemical dissociations at the surfaces. We found that charge transfer from the polysulfides to the WX$_2$ is a key factor that dictates the binding strength. Overall, these findings provide insight into a fundamental understanding of polysulfides chemistry on the WX$_2$ surfaces to discern the experimentally reported superior performance of WX$_2$ towards polysulfides suppression.

## 2. Calculation methods

We performed spin polarized plane wave DFT calculations using the VASP software package.[44] We used Projector Augmented Wave (PAW) pseudopotential to study the electron-ion interaction and Perdew-Bruke-Ernzerhof (PBE) functional with the Generalized Gradient Approximations (GGA) approximation to obtain the electron-electron exchange correlations.[45] We used Grimme's DFT-D2 method to account for the van der Waal interactions.[46] This treatment is essential to accurately evaluate the binding strength of the polysulfides with the AMs. The kinetic cutoff energy for the plane wave basis calculations was chosen to be 520 eV. The convergence criterion for self-consistent field calculations was used as 10$^{-4}$ eV and atom positions were relaxed up to force tolerance of 0.025 eV/Å. We used 5 x 5 x 1 supercell of the unit cells of WS$_2$ and WSe$_2$ consists of 75 atoms. A vacuum spacing of 25 Å was applied to the normal direction in order to eliminate the interactions across the periodic boundary. The Brillouin zone was sampled using a 5 x 5 x 1 and 11 x

11 x 1 *k-point* mesh generated by the Monkhorst-Pack scheme for relaxation and density of states (DOS) calculations, respectively. The Brillouin zone integration was accomplished by the tetrahedron method with Blöchl corrections using a broadening width of 0.05 eV. The charger transfer during the LiPSs adsorption on the AM was studied using the Bader charge analysis[47] and the charge differences were calculated using the following expression

$$\rho_b = \rho_{adsorbed\,state} - (\rho_{adsorbent} + \rho_{AM})$$

Where $\rho_{adsorbed\,state}$, $\rho_{adsorbent}$, and $\rho_{AM}$ indicates the charge density of the AM with adsorbed LiPS (WX$_2$-LiPS), isolated LiPS, and the AM, respectively. The charge density differences were visualized using the VESTA code.[47]

### 3. Results and Discussions

The lithium ions during cell discharge released from the anode diffuse through the electrolyte and reduces the sulfur cathode to produce soluble intermediate lithium polysulfides such as Li$_2$S$_8$, Li$_2$S$_6$, Li$_2$S$_4$, and finally insoluble Li$_2$S$_2$ and Li$_2$S. The anchoring materials are required to provide containment for the polysulfides within the cathode material. In order to evaluate the effectiveness of an AM, it is essential to understand the binding behavior between the AM and the polysulfides. Herein, we performed a detailed analysis of the binding of various polysulfides with both WS$_2$ and WSe$_2$ host materials.

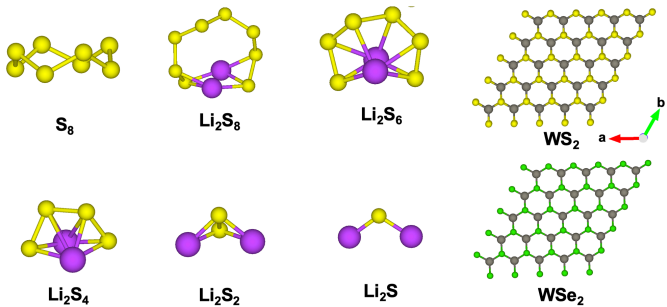

Figure 1. The optimized geometries of S$_8$, LiPSs, and top view of WS$_2$ and WSe$_2$ structures. Color codes: Yellow: Sulfur, Purple: Lithium, Green: Selenium, and Gray: Tungsten.

First, we optimize the elemental sulfur, the major LiPSs and AM to study their chemical structures. The puckered ring structure of elemental sulfur (S$_8$) has a D$_{4d}$ symmetry with the S-S bond length of about 2.06 Å. The higher order polysulfides such as Li$_2$S$_8$, Li$_2$S$_6$, and Li$_2$S$_4$ exhibit C$_2$ symmetry and the shortest Li-S and S-S bonds are 2.40 Å and 2.07 Å, respectively. In contrast, the insoluble lower order Li$_2$S$_2$ and Li$_2$S have C$_{2v}$ symmetry. The obtained bond lengths and the listed symmetries are in accordance with the previous studies, which establishes the accuracy of our calculations.[48] The optimized structures of the S$_8$, LiPSs, WS$_2$, and WSe$_2$ are shown in Figure 1.

Next, we perform structural relaxation simulations to investigate the adsorption behavior of LiPSs on the AM. The optimized configurations of the LiPSs anchored on the WS$_2$ and WSe$_2$ are shown in Figure 2. One can see that the cyclooctane sulfur (S$_8$) has a different optimized structure when compared to the other polysulfides. The geometric orientation of the absorbed sulfur and LiPSs are analogous in both WS$_2$ and WSe$_2$. The structural deformation of the LiPS species is apparent as shown in Figure 2 whereas the AMs retain the same structure even after adsorption. The Li atoms in LiPSs remain close to the AMs similar to the graphene, MoS$_2$,[22] TiS$_2$,[22] blue phosphorene,[49] and black phosphorene,[39] but in contrary to the borophene[43] where S atoms are closer to the AMs. This difference in the stereo configurations of the adsorbed LiPSs can be attributed to the differences in the net electronegativity values between the AM and the S in the LiPSs.

In order to understand the anchoring strength of the host materials, the binding energies were computed. The adsorption energies of the polysulfides with the AM was calculated using the formula

$$E_{ads} = E_{Li_2S_n} + E_{AM} - E_{Li_2S_n+AM}$$

Where $E_{Li_2S_n}$, E$_{AM}$, and $E_{Li_2S_n+AM}$, denotes the energy of isolated polysulfides, AM, and the polysulfides adsorbed system, respectively. The calculated adsorption energies for WX$_2$ (X = S, Se) are presented in Table 1. The positive values for the adsorption energies indicate that the adsorption of polysulfides are energetically favourable. The Table 1 shows that the binding energies for WS$_2$ adsorbed structures are 0.62 eV for S$_8$, 0.67 eV for Li$_2$S$_8$, 0.56 eV for Li$_2$S$_6$, 0.89 eV for Li$_2$S$_4$, 1.08 eV for Li$_2$S$_2$, and 1.22 eV for Li$_2$S.

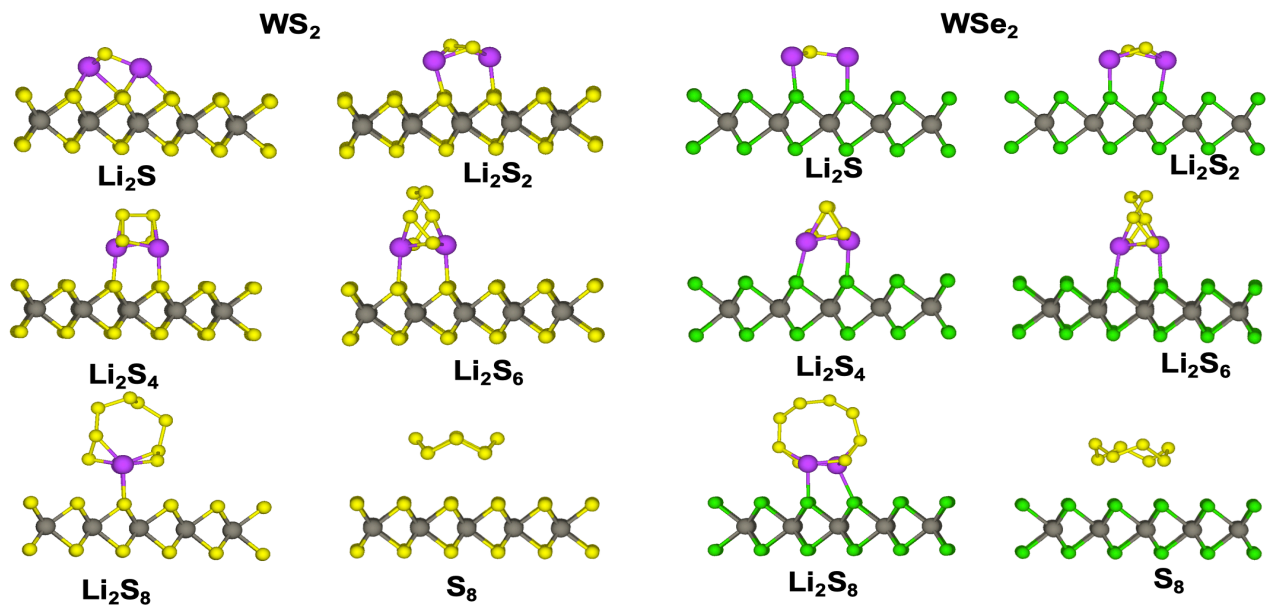

Figure 2. The optimized geometries of $S_8$ and LiPSs adsorbed on $WS_2$ and $WSe_2$. The LiPSs form chemical bonding with $WX_2$ while $S_8$ interacts primarily through vdW.

For $WSe_2$ adsorbed structures, the binding energies are in the order of 0.66 eV for $S_8$, 0.77 eV for $Li_2S_8$, 0.60 eV for $Li_2S_6$, 0.91 eV for $Li_2S_4$, 1.04 eV for $Li_2S_2$, and 1.07 eV for $Li_2S$. We found that the $S_8$ binding energies on $WS_2$ and $WSe_2$ are lower compared to pristine graphene (0.75 eV), and various metal oxides such as $V_2O_5$ (0.80 eV) and $MoO_3$ (0.78 eV), however, comparable to $MoS_2$ (0.67 eV).[22] The lower binding energy of $S_8$ can be attributed to the absence of lithium atoms which causes charge transfer induced chemical bonds, while sulfur only interact with the host material via nonbonded van der Waal interaction. Zhang et al.[22] has studied the binding strength for numerous oxides ($MoO_3$, $V_2O_5$), sulfides ($TiS_2$, $NbS_2$, $MoS_2$, $VS_2$), chlorides ($TiCl_2$, $ZrCl_2$), and concluded that all the AMs can be grouped into three categories such as strong, moderate, and weak AMs. According to this classification, the binding strengths of LiPSs on the $WS_2$ and $WSe_2$ indicates these materials can be categorized as moderate strength AMs. The characteristic of the binding strength of the polar AMs can be partly ascribed to the differences in the electronegativity ($\Delta\chi$) of the constituent elements. The relatively smaller $\Delta\chi$ of transition metal (TM) sulfides/selenides results in weaker binding compared to the TM oxides with the larger $\Delta\chi$. For example, our calculated LiPSs binding energies on $WX_2$ are approximately four times smaller than the $V_2O_5$.[22] Our calculations reveal that the polysulfides formed with the increased lithiation exhibit higher binding energies except a slight anomalous binding behaviour of $Li_2S_6$. This observation is consistent with the previous reports on $Li_2S_6$ adsorption on various transition metal disulfides.[22] Interestingly, the trend in the binding energies with the increased lithiation is opposite to that of the graphene, as in graphene during lithiation the binding strength decreases. Overall, we found that the anchoring behaviour of $WX_2$ is similar to $MoS_2$,[22] and the behaviour is expected because of the chemical similarity between the Mo and W.

To further investigate the LiPSs bonding with the AMs, we calculate the smallest bond distance between the LiPSs, and the AMs and the data are presented in Table 1. The shortest bond distance between the adsorbed $S_8$ molecule and the $WS_2$ and $WSe_2$ are found as 3.53 Å and 3.64Å, respectively. The $S_8$ maintained its puckered ring structure even after adsorption and the longer distance implies the absence of any chemical bonds. The obtained bond distance for $S_8$ cluster is similar to that of other AMs such as blue phosphorene,[49] black phosphorene,[50] N-doped and amorphous graphene,[22,42] $Ti_2Co_2$,[51] and $C_3B$.[52] In LiPSs adsorbed structures, the bond distance between the Li in LiPSs and the S and Se in AMs decreases with the decrease in sulfur content in LiPSs. The trend in the increase of the shortest distance with the increment of S concentration in the LiPSs is correlated with the corresponding binding energies (see Table 1). We found that the adsorption of the LiPSs on the AMs causes an increase in the intramolecular Li-S bond length and this increase manifests stronger interactions of LiPSs with the AMs and the similar effect has been reflected in the binding energy values as well.

Table 1. The shortest distance, the minimum distance between the AMs and the LiPS, and the charge transfer of various LiPSs adsorbed WS$_2$ and WSe$_2$

| LiPS | WS$_2$ | | | | WSe$_2$ | | | |
|---|---|---|---|---|---|---|---|---|
| | Δd$_{Li-S}$ | d$_{Li-AM}$ | E$_b$ (eV) | q(e) | Δd Li-S | d$_{Li-AM}$ | E$_b$ (eV) | q (e) |
| S$_8$ | - | 3.53 | 0.62 | 0.007 | - | 3.64 | 0.66 | -0.010 |
| Li$_2$S$_8$ | 0.020 | 2.71 | 0.67 | 0.103 | 0.040 | 2.85 | 0.77 | 0.060 |
| Li$_2$S$_6$ | 0.003 | 2.69 | 0.56 | 0.110 | 0.000 | 2.83 | 0.60 | 0.063 |
| Li$_2$S$_4$ | 0.020 | 2.59 | 0.89 | 0.130 | 0.030 | 2.71 | 0.91 | 0.090 |
| Li$_2$S$_2$ | 0.060 | 2.45 | 1.08 | 0.331 | 0.040 | 2.58 | 1.04 | 0.260 |
| Li$_2$S | 0.100 | 2.52 | 1.22 | 0.567 | 0.050 | 2.60 | 1.07 | 0.349 |

The characteristic of the bonding behaviour between the LiPSs and the AMs were further probed using the Bader charge analysis and charge density difference. The Bader charge analysis provides an estimation of the amount of charge transfer that occurred between the LiPSs and AMs. The data extracted from the analysis are presented in Table 1. The calculated charge transfer values for WS$_2$ adsorbed structures are 0.007e| for S$_8$, 0.103 |e| for Li$_2$S$_8$, 0.11 |e| for Li$_2$S$_6$, 0.13|e| for Li$_2$S$_4$, 0.331|e| for Li$_2$S$_2$, and 0.567 |e| for Li$_2$S. For WSe$_2$, the values are -0.01 |e| for S$_8$, 0.060|e| for Li$_2$S$_8$, 0.063 |e| for Li$_2$S$_6$, 0.090|e| for Li$_2$S$_4$, 0.260 |e| for Li$_2$S$_2$, and 0.349 |e| for Li$_2$S. The positive values for the charge transfer indicate that the charge is transferred from the polysulfides to the AMs. The weaker interactions between the unlithiated S$_8$ and the AMs are manifested through the insignificantly smaller values of charge transfer. The only interaction between the two species arises due to the nonbonded van der Waal interactions.

With the increased lithiation, the higher quantity of charge is transferred from the polysulfides to the AMs and the charge transfer induces the formation of chemical bonds, that is, the interactions between the positively charged Li in LiPS species and negatively charged S or Se in WX$_2$ results in the Li-S/Se bonds. The charge bereaved from the LiPS structures responsible for the elongation and weakening of Li-S bonds after adsorption. The observed slightly anomalous binding strength of Li$_2$S$_6$ relative to the order of the polysulfides can be partly ascribed to the lower quantity of charge transfer for this case.

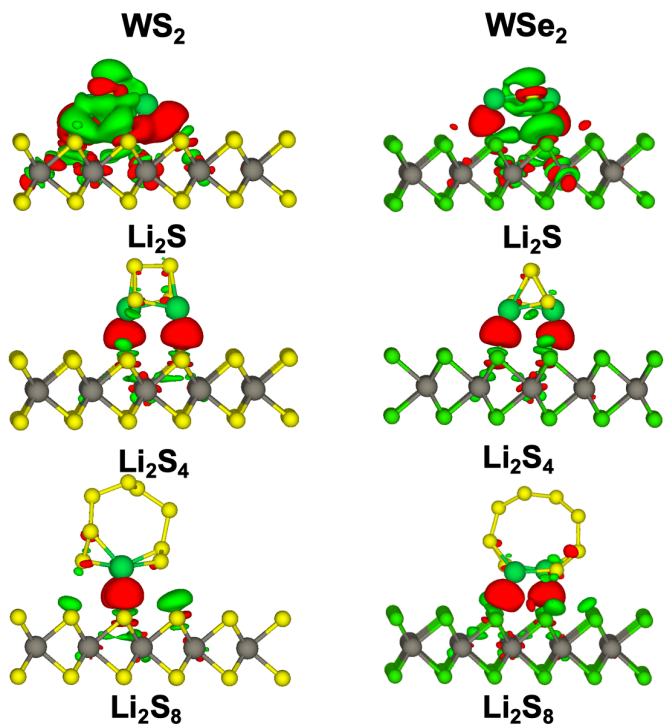

Figure 3. Charge density differences for Li$_2$S, Li$_2$S$_4$, and Li$_2$S$_8$ on WS$_2$ and WSe$_2$. The isosurface level is set to 0.001 e Å$^{-3}$. The green and red indicate depletion and gain of charges, respectively.

The anchoring effect of the materials can be increased with the increase in the chemical interaction arising due to the higher charge transfer. The charge transfer from the LiPSs to the AMs can be further visualized from the charge density difference analysis. The charge density difference was performed for $Li_2S$, $Li_2S_4$, and $Li_2S_8$ adsorbed AMs and the data are shown in Figure 3. The figures illustrate the depletion of charges in the LiPSs and the correspondingly charge localization in the AMs. The quantity of the charge redistribution for various LiPSs also corroborates the charge transfer data.

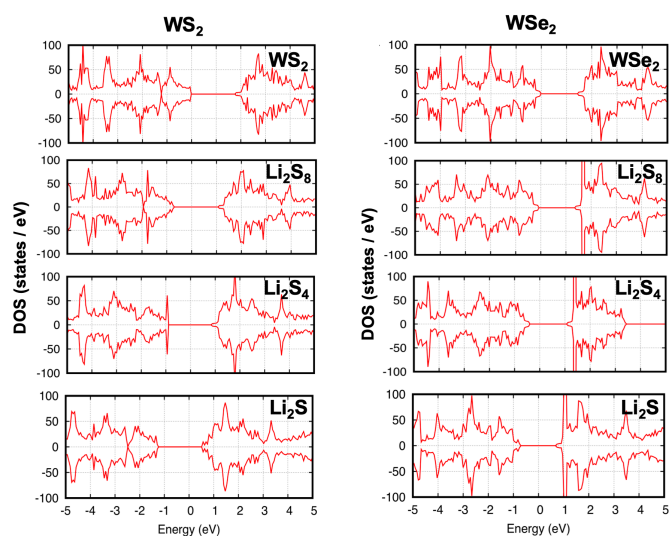

Figure 4. The total density of states of pristine, $Li_2S$, $Li_2S_4$, and $Li_2S_8$ adsorbed $WS_2$ and $WSe_2$. The lithiation resulted in the downshifting of the DOS.

In order to understand the electronic behavior of the AMs, the total density of states (DOS) was calculated for both pristine and adsorbed structures. The DOS of pristine $WX_2$, $Li_2S$, $Li_2S_4$, and $Li_2S_8$ adsorbed structures (displayed in Figure 4) are analyzed to elucidate how LiPSs adsorption changes the electronic structures of the AMs. The band gap of the pristine $WS_2$ and $WSe_2$ are predicted as 1.77 eV and 1.46 eV which compares well with the previously reported bandgaps of these materials.[53–55] We observe a similar trend in the changes of electronic structures of both $WS_2$ and $WSe_2$ with the adsorption of various LiPSs. The electron transfer due to the adsorption of LiPS species to the AMs results in a downshift of the DOS, that is, the lowering of both the conduction band maxima (CBM) and valence band minima (VBM) relative to the Fermi level. However, the predicted band gaps of the LiPSs absorbed structures are comparable to the pristine $WX_2$. The increase in the charge transfer with the lower order polysulfides causes more pronounced downshifting of the DOS. The electrons from the LiPS species are found to be filled in the 5-d state of W. Overall, the DOS calculations depict that adsorption of the LiPSs onto the AMs has a relatively insignificant influence on the band gap characteristics of the $WX_2$ materials and the materials behaves as semiconductors.

## 4. Conclusion

In summary, we used density functional theory (DFT) calculations to understand the anchoring effect of the tungsten dichalcogenides such as $WS_2$ and $WSe_2$. The binding energy values are almost similar for both the materials and the results reveal that they could exhibit not too strong but adequate binding characteristics. The binding energy values lie between 0.56 to 1.22 eV for both the materials and binding strength depends primarily on the charge transfer from the polysulfides to the AMs. The LiPSs retain their chemical structures after adsorption which is indicative of good cyclability during charging and discharging. The analysis of Bader charge, differential charge density, and DOS were carried out to understand further about the LiPSs binding mechanism in these materials. The charge analysis indicates that the charge transfers from the polysulfide species to the AMs and we visualized the electron localization using the differential charge density analysis. The total DOS calculations reveal both the AMs manifest a downshift from the fermi level after adsorption of the polysulfides and the materials preserve their semiconducting behaviour. Finally, the outcomes of this work would potentially elicit further computational and experimental studies on various other 2D anchoring materials to develop cathodes for high-performance Li-S

**Acknowledgments**

M.M.I acknowledges the start-up funds from Wayne State University and the support of the Extreme Science and Engineering Discovery Environment (XSEDE)[56] for providing the computational facilities (Start up Allocation – DMR190089).